# Around Dot-depth One[*]

Manfred Kufleitner     Alexander Lauser

University of Stuttgart, FMI
{kufleitner, lauser}@fmi.uni-stuttgart.de

March 7, 2011

**Abstract**

The dot-depth hierarchy is a classification of star-free languages. It is related to the quantifier alternation hierarchy of first-order logic over finite words. We consider fragments of languages with dot-depth 1/2 and dot-depth 1 obtained by prohibiting the specification of prefixes or suffixes. As it turns out, these language classes are in one-to-one correspondence with fragments of existential first-order logic without min- or max-predicate. For all fragments, we obtain effective algebraic characterizations. Moreover, we give new combinatorial proofs for the decidability of the membership problem for dot-depth 1/2 and dot-depth 1.

## 1 Introduction

The dot-depth hierarchy $\mathcal{B}_n$ for $n \in \mathbb{N} + \{1/2, 1\}$ has been introduced by Cohen and Brzozowski [2]. A very similar hierarchy is the Straubing-Thérien hierarchy $\mathcal{L}_n$, see [16, 18]. Both hierarchies are strict [1] and they are exhausting the class of star-free languages. A classical result of McNaughton and Papert is that a language is star-free if and only if it is definable in first-order logic [9]. Thomas [20] has tightened this result by showing that there is a one-to-one correspondence between the dot-depth hierarchy (and also between the Straubing-Thérien hierarchy) and the quantifier alternation hierarchy of first-order logic. More precisely, the dot-depth hierarchy is related to the quantifier alternation hierarchy over the signature $[<, +1, \min, \max]$, whereas the Straubing-Thérien hierarchy corresponds to the quantifier alternation hierarchy over the signature $[<]$.

Schützenberger has shown that a language is star-free if and only if its syntactic semigroup is aperiodic [13]. The latter property is effectively decidable. Together with the result of McNaughton and Papert, this yields a decision procedure for definability in first-order logic. Effectively determining the level of a language in the dot-depth hierarchy or equivalently, in the quantifier alternation hierarchy of first-order logic, is one of the most challenging open problems in automata theory. For $n \in \mathbb{N}$, Straubing has shown that membership in $\mathcal{B}_n$ is decidable if and only if membership in $\mathcal{L}_n$ is decidable [17]. This result has been extended to the half-levels by Pin and Weil [12]. Simon has shown that the class of piecewise testable languages $\mathcal{L}_1$ is decidable [14]. Later, Knast [6] gave an effective algebraic characterization of $\mathcal{B}_1$. Decidability of $\mathcal{L}_{1/2}$ was shown by Pin [10], and the levels $\mathcal{B}_{1/2}$ and $\mathcal{L}_{3/2}$ are decidable by a result of Pin and Weil [11]. The most recent decidability result is for $\mathcal{B}_{3/2}$ due to Glaßer and Schmitz [4]. To date, no other levels are known to be decidable.

[*]Supported by the German Research Foundation (DFG), project DI 435/5-1.

In this paper, we focus on subclasses of $\mathcal{B}_{1/2}$ and $\mathcal{B}_1$. For both $\mathcal{B}_{1/2}$ and $\mathcal{B}_1$ we give new proofs for their effective algebraic characterizations. The proof of Pin and Weil [11] for $\mathcal{B}_{1/2}$ is based on factorization forests [15], and the proof of Knast [6] as well as the simplified version of Thérien [19] for $\mathcal{B}_1$ are based on a generalization of finite monoids, so-called *finite categories* [21]. Our proofs are more combinatorial than algebraic. The proof for $\mathcal{B}_1$ is a generalization of Klíma's proof [5] for $\mathcal{L}_1$. The main advantage of our proofs for $\mathcal{B}_{1/2}$ and $\mathcal{B}_1$ over previous ones is that the constants involved in finding language descriptions for given algebraic objects are more explicit (and therefore smaller).

Our main original contributions are effective algebraic characterizations of fragments of existential first-order logic over the signatures $[<, +1, \min]$ without max-predicate, $[<, +1, \max]$ without min, and $[<, +1]$ without min and max. These fragments also admit language characterizations in terms of subclasses of $\mathcal{B}_{1/2}$ and $\mathcal{B}_1$. The corresponding language classes are obtained by prohibiting the specification of prefixes or suffixes. In contrast to $\mathcal{B}_{1/2}$ and $\mathcal{B}_1$, the resulting subclasses do not form (positive) varieties of languages, but they still can be described using so-called lattice equations [3]. Moreover, there is a tight connection with Cantor topologies over finite words [7]. A more detailed overview of our results can be found in Section 7.

## 2 Preliminaries

**Words and Languages** Let $\Gamma$ be a finite non-empty alphabet. The set of finite words is $\Gamma^*$. By $\varepsilon$ we denote the empty word and $\Gamma^+ = \Gamma^* \setminus \{\varepsilon\}$ is the set of finite non-empty words. The length of a word $u \in \Gamma^*$ is $|u|$ and its alphabet is $\mathrm{alph}(u) = \{a \in \Gamma \mid u \in \Gamma^* a \Gamma^*\}$. Similarly, $\mathrm{alph}_k(u) = \{v \in \Gamma^k \mid u \in \Gamma^* v \Gamma^*\}$ is the set of all factors of $u$ of length $k$. A word $v \in \Gamma^*$ is a prefix (resp. suffix, resp. factor) of $u$ if $u \in v\Gamma^*$ (resp. $u \in \Gamma^* v$, resp. $u \in \Gamma^* v \Gamma^*$). We write $v \leq_p u$ if $v$ is a prefix of $u$ and $v <_p u$ if $v$ is a proper prefix of $u$. A *quotient* of $L \subseteq \Gamma^+$ is a language of the form $u^{-1}L = \{v \in \Gamma^+ \mid uv \in L\}$ or $Lu^{-1} = \{v \in \Gamma^+ \mid vu \in L\}$ for $u \in \Gamma^*$. A language $L$ is a *monomial* (of degree $m$) if $L = w_1 \Gamma^* w_2 \cdots \Gamma^* w_n$ or $L = w_1 \Gamma^+ w_2 \cdots \Gamma^+ w_n$ for some $n \geq 0$ and $w_1, \ldots, w_n \in \Gamma^*$ (with $|w_1 \cdots w_n| \leq m$). A language has *dot-depth one* if it is a Boolean combination of monomials. Throughout this paper, Boolean operations are complementation, *finite* union, and *finite* intersection. Positive Boolean operations are finite union and finite intersection.

**First-order Logic over Words** We consider the first-order logic $\mathrm{FO} = \mathrm{FO}[<, +1, \min, \max]$ over finite words. We view words as sequences of labeled positions which are linearly ordered by $<$. Variables are interpreted as positions of a word. For variables $x, y$ we have the following atomic formulas: $x < y$ says that $x$ is a position smaller than $y$; and $x = y + 1$ is true if $x$ is the immediate successor of $y$; the formula $\min(x)$ (resp. $\max(x)$) holds if $x$ is the first (resp. last) position. Moreover, we always assume that we have an atomic formula $\top$ (for *true*), equality of positions $x = y$, and a predicate $\lambda(x) = a$ specifying that position $x$ is labeled by $a \in \Gamma$. Formulas can be composed using Boolean operations, existential quantification, and universal quantification. The semantics is as usual. A sentence is a formula without free variables. For a sentence $\varphi$ of FO we write $u \models \varphi$ if $u$ is a model of $\varphi$ and the language defined by $\varphi$ is $L(\varphi) = \{u \in \Gamma^+ \mid u \models \varphi\}$.

The fragment $\Sigma_1$ consists of all FO-formulas in prenex normal form with only one block of quantifiers and these quantifiers are existential. Let $\mathcal{C} \subseteq \{<, +1, \min, \max\}$. By $\Sigma_1[\mathcal{C}]$ we denote the class of formulas in $\Sigma_1$ which only use predicates in $\mathcal{C}$, equality, and the label predicate. The fragment $\mathbb{B}\Sigma_1[\mathcal{C}]$ comprises all Boolean combinations of formulas in $\Sigma_1[\mathcal{C}]$.

**Finite Semigroups and Recognizable Languages** Let $S$ be a semigroup. An element $x \in S$ is *idempotent* if $x^2 = x$. The set of idempotents is denoted by $E(S) = \{e \in S \mid e^2 = e\}$. For every finite semigroup $S$ there exists a number $\omega \geq 1$ such that for every $x \in S$, the power $x^\omega$ is the unique idempotent element generated by $x$. Frequently, we consider words $u, v \in S^*$ where the alphabet is a semigroup. We write "$u = v$ in $S$" if either $u = \varepsilon = v$ or $u, v \in S^+$ evaluate to the same element of $S$.

**Lemma 1.** *Let $S$ be a finite semigroup. For every word $u \in S^+$ with length $|u| = |S|$ there exists a non-empty prefix $p$ of $u$ and an idempotent $e \in E(S)$ such that $pe = p$ in $S$.*

*Proof.* Let $a \in S$ be arbitrary and let $p_1 <_p \cdots <_p p_{|S|} <_p p_{|S|+1} = ua$ be the non-empty prefixes of $ua$. By the pigeonhole principle, there exist $1 \leq i < j \leq |S| + 1$ such that $p_i = p_j$ in $S$. In particular, $i \leq |S|$ and $p_i$ is a prefix of $u$. Let $p_i q = p_j$ for $q \in S^+$. We set $e = q^\omega$ to be the idempotent element generated by $q$. Now, $pe = p$ in $S$ for $p = p_i$. □

Green's relations are an important tool in the study of semigroups. They are defined as follows. Let $x \leq_\mathcal{R} y$ (resp. $x \leq_\mathcal{L} y$, resp. $x \leq_\mathcal{J} y$) if there exist $s, t \in S \dot\cup \{1\}$ such that $x = yt$ (resp. $x = sy$, resp. $x = syt$). Let $x \mathcal{R} y$ (resp. $x \mathcal{L} y$, resp. $x \mathcal{J} y$) if $x \leq_\mathcal{R} y$ and $y \leq_\mathcal{R} x$ (resp. $x \leq_\mathcal{L} y$ and $y \leq_\mathcal{L} x$, resp $x \leq_\mathcal{J} y$ and $y \leq_\mathcal{J} x$). Here, $S \dot\cup \{1\}$ is the monoid obtained by adding a new neutral element 1 to the semigroup $S$. The relations $\leq_\mathcal{R}, \leq_\mathcal{L}$, and $\leq_\mathcal{J}$ are preorders on $S$; and $\mathcal{R}, \mathcal{L}$, and $\mathcal{J}$ form equivalence relations.

Let $\leq$ be a preorder on $S$. A set $P \subseteq S$ is a $\leq$-*order ideal* if $x \leq y \in P$ implies $x \in P$. The order ideal generated by some subset $P \subseteq S$ is $\downarrow P = \{x \in S \mid x \leq y \text{ for some } y \in P\}$. An *ordered semigroup* $S$ is equipped with a compatible partial order $\leq$, i.e., if $p \leq q$ and $s \leq t$, then $ps \leq qt$. Every semigroup is an ordered semigroup with equality as partial order. A language $L \subseteq \Gamma^+$ is *recognized* by an ordered semigroup $S$ if there exists a homomorphism $h : \Gamma^+ \to S$ such that $L = h^{-1}(P)$ for some $\leq$-order ideal $P$. If the order of $S$ is equality, then we obtain the usual notion of recognition. For a language $L \subseteq \Gamma^+$ the *syntactic preorder* $\leq_L$ over $\Gamma^+$ is given by $x \leq_L y$ if $uyv \in L \Rightarrow uxv \in L$ for all $u, v \in \Gamma^*$. The syntactic congruence $\equiv_L$ is defined by $x \equiv_L y$ if both $x \leq_L y$ and $y \leq_L x$. The equivalence classes of the syntactic congruence equipped with the canonical composition constitutes the *syntactic semigroup* $\mathrm{Synt}(L)$ and the preorder $\leq_L$ of $\Gamma^+$ becomes a compatible partial order for $\mathrm{Synt}(L)$. The syntactic semigroup of $L$ is finite if and only if $L$ is regular and moreover, $L$ is recognized by its syntactic semigroup. By $[\![x^\omega y x^\omega \leq x^\omega]\!]$ we denote the class of finite ordered semigroups $S$ such that $x^\omega y x^\omega \leq x^\omega$ for all elements $x, y \in S$. We let $\mathbf{B_1}$ be the class of finite semigroups $S$ such that $(exfy)^\omega exf(tesf)^\omega = (exfy)^\omega esf(tesf)^\omega$ for all idempotents $e, f \in E(S)$ and all elements $x, y, t, s \in S$. Let $\mathbf{LR}$ be the class of finite semigroups $S$ such that $(exeye)^\omega exe = (exeye)^\omega$ for all idempotents $e \in E(S)$ and all elements $x, y \in S$. We have the following inclusions among these classes of semigroups.

**Lemma 2.** *We have $[\![x^\omega y x^\omega \leq x^\omega]\!] \subseteq \mathbf{B_1} \subseteq \mathbf{LR}$.*

*Proof.* For a semigroup $S \in [\![x^\omega y x^\omega \leq x^\omega]\!]$ we have $f \geq fy(exfy)^{\omega-1}esf$ for all $x, y, s \in S$ and all idempotents $e, f \in S$. Hence $(exfy)^\omega exf(tesf)^\omega \geq (exfy)^\omega ex(fy(exfy)^{\omega-1}esf)(tesf)^\omega = (exfy)^\omega esf(tesf)^\omega$. By symmetry $(exfy)^\omega exf(tesf)^\omega \leq (exfy)^\omega esf(tesf)^\omega$ proving the first inclusion.

We have $(exey)^\omega exe = (exey)^\omega exe(eeee)^\omega$ for all $x$ and all idempotents $e$ and for a semigroup in $\mathbf{B_1}$ this is equal to $(exey)^\omega eee(eeee)^\omega = (exeye)^\omega$. This shows the second inclusion. $\square$

## 3 Dot-depth 1/2

A language $L \subseteq \Gamma^+$ has dot-depth 1/2 if it is a positive Boolean combination of monomials $w_1 \Gamma^* w_2 \cdots \Gamma^* w_n$ with $w_i \in \Gamma^*$. By a result of Thomas [20], a language has dot-depth 1/2 if and only if it is definable in existential first-order logic $\Sigma_1[<, +1, \min, \max]$. Pin and Weil [11] have shown that $L$ has dot-depth 1/2 if and only if $\mathrm{Synt}(L) \in [\![x^\omega y x^\omega \leq x^\omega]\!]$. In this section, we give a new proof of these equivalences. The main step in the proof is to show that if $L$ is recognized by some homomorphism $h : \Gamma^+ \to S \in [\![x^\omega y x^\omega \leq x^\omega]\!]$, then $L$ is a union of monomials $w_1 \Gamma^* w_2 \cdots \Gamma^* w_n$. The main advantage of our proof is that the degree $|w_1 \cdots w_n|$ is polynomially bounded (Proposition 9), whereas in the proof of Pin and Weil, the bound is exponential.

**Theorem 3** (Thomas [20], Pin/Weil [11])**.** *Let $L \subseteq \Gamma^+$. The following assertions are equivalent:*

1. *$L$ is definable in $\Sigma_1[<, +1, \min, \max]$.*
2. *$L$ is a finite union of monomials $w_1 \Gamma^* w_2 \cdots \Gamma^* w_n$.*
3. *$L$ is a positive Boolean combination of monomials $w_1 \Gamma^* w_2 \cdots \Gamma^* w_n$.*
4. *$\mathrm{Synt}(L) \in [\![x^\omega y x^\omega \leq x^\omega]\!]$.*
5. *There exists a homomorphism $h : \Gamma^+ \to S$ with $S \in [\![x^\omega y x^\omega \leq x^\omega]\!]$ such that $L = h^{-1}(P)$ for some $\leq$-order ideal $P$.*

In the remainder of this section we prove the above theorem.

**Lemma 4.** *Let $n \geq 0$, and let $w_1, \ldots, w_n \in \Gamma^*$.*
1. *The monomial $w_1 \Gamma^* w_2 \cdots \Gamma^* w_n$ is definable in $\Sigma_1[<, +1, \min, \max]$.*
2. *The monomial $w_1 \Gamma^* w_2 \cdots \Gamma^* w_n \Gamma^*$ is definable in $\Sigma_1[<, +1, \min]$.*
3. *The monomial $\Gamma^* w_1 \Gamma^* w_2 \cdots \Gamma^* w_n \Gamma^*$ is definable in $\Sigma_1[<, +1]$.*

*Proof.* The proof is straightforward. For variable vectors $\underline{x} = (x_1, \ldots, x_\ell)$ and $\underline{y} = (y_1, \ldots, y_m)$ we use the shortcuts $\exists \underline{x}$ for $\exists x_1 \cdots \exists x_\ell$, and $\min(\underline{x})$ for $\min(x_1)$ and $\max(\underline{x})$ for $\max(x_\ell)$, and $\underline{x} < \underline{y}$ means $x_\ell < y_1$. Moreover, $\lambda(\underline{x}) = a_1 \cdots a_k$ is a shortcut for

$$\bigwedge_{1 \leq j \leq k} \lambda(x_j) = a_j \;\wedge\; \bigwedge_{1 \leq j < k} x_{j+1} = x_j + 1.$$

Let $L = \Gamma^* w_1 \Gamma^* w_2 \cdots \Gamma^* w_n \Gamma^*$. We introduce variable vectors $\underline{x}_i = (x_{i,1}, \ldots, x_{i,|w_i|})$ for every $i \in \{1, \ldots, n\}$. Then, $L$ is defined by the following sentence $\varphi$:

$$\exists \underline{x}_1 \cdots \exists \underline{x}_n \colon \bigwedge_{1 \leq i \leq n} \lambda(\underline{x}_i) = w_i \wedge \bigwedge_{1 \leq i < n} \underline{x}_i < \underline{x}_{i+1}$$

The first term of the conjunction ensures that each $\underline{x}_i$ corresponds to a factor $w_i$, whereas the second term ensures that the factors $w_i$ occur in the correct order. The sentence for $w_1 \Gamma^* w_2 \cdots \Gamma^* w_n \Gamma^*$ is $\varphi \wedge \min(\underline{x}_1)$ and the sentence for $w_1 \Gamma^* w_2 \cdots \Gamma^* w_n$ is $\varphi \wedge \min(\underline{x}_1) \wedge \max(\underline{x}_n)$. □

**Lemma 5.** *Let $L \subseteq \Gamma^+$ be definable by a sentence $\varphi \in \Sigma_1[<, +1, \min, \max]$ with $m$ variables. Then $L$ is a finite union of monomials $w_1 \Gamma^* w_2 \cdots \Gamma^* w_n$ of degree at most $m$.*

*Proof.* Let $\varphi = \exists x_1 \cdots x_m \colon \psi$ for a propositional formula $\psi$. Suppose $(u, x_1, \ldots, x_m) \models \psi$ for positions $x_i$ of $u$. We say that a position $j$ of $u$ is *marked* if $j = x_i$ for some $i$. In order to avoid case distinctions we can introduce two new variables such that the first and the last position of $u$ are marked. Let $u = w_1 u_1 w_2 \cdots u_{n-1} w_n$ for $u_i \in \Gamma^+$ such that the factors $w_i$ consist of the marked positions. Now, $P_u = w_1 \Gamma^+ w_2 \cdots \Gamma^+ w_n$ is a monomial of degree $|w_1 \cdots w_n| \leq m$ with $u \in P_u$. Moreover, $P_u \subseteq L(\varphi)$ since the satisfying assignment of $u$ can be adapted to all $v \in P_u$. It follows $L(\varphi) = \bigcup_{u \models \varphi} P_u$ and this union is finite since there are only finitely many monomials of degree at most $m$. □

**Lemma 6.** *Let $L \subseteq \Gamma^+$ be a finite union of monomials $w_1 \Gamma^* w_2 \cdots \Gamma^* w_n$. Then $\mathrm{Synt}(L) \in [\![x^\omega y x^\omega \leq x^\omega]\!]$.*

*Proof.* Let $P = w_1 \Gamma^* w_2 \cdots \Gamma^* w_n$ and let $u, x, y, v \in \Gamma^+$ and choose $m$ such that $|x^m| > |w_1 \cdots w_n|$. Suppose $u x^m v \in P$. Let $i$ be maximal such that $u x^m \in w_1 \Gamma^* w_2 \cdots \Gamma^* w_i \Gamma^* = Q_i$ and let $j$ be minimal such that $x^m v \in \Gamma^* w_j \cdots \Gamma^* w_n = R_j$. By the choice of $m$ we have $j \leq i + 1$. Hence, $u x^m y x^m v \in Q_i R_j \subseteq P$. □

**Lemma 7.** *Let $S$ be a finite semigroup. For every $w \in S^+$ there exists a factorization $w = x_1 w_1 y_1 \cdots x_m w_m y_m s$ with*
1. *$w_i, s \in S^*$, $x_i, y_i \in S^+$, $|y_i| \leq |S|$,*
2. *$0 \leq m \leq |S|$ and $|x_1 y_1 \cdots x_m y_m s| < 2|S|^2 + |S|$,*
3. *$\forall i \in \{1, \ldots, m\} \ \exists e_i \in E(S) \colon x_i = x_i e_i$ in $S$ and $y_i = y_i e_i$ in $S$.*

*Proof.* For $w \in S^*$, let $E(w)$ be the set all $e \in E(S)$ such that there exists a factor $x \in S^+$ of $w$ with $|x| \leq |S|$ and $xe = x$ in $S$. We prove the existence of the factorization by induction on $|E(w)|$ with the stronger assertions that $m \leq |E(w)|$ and $|x_1 y_1 \cdots x_m y_m s| < 2|S| |E(w)| + |S|$ instead of condition "2". Suppose $|E(w)| = 0$. By Lemma 1 we have $|w| < |S|$. Hence, we can choose $m = 0$ and $s = w$.

If $|E(w)| \geq 1$, then Lemma 1 yields a non-empty prefix $x$ of $w$ with $|x| \leq |S|$ such that $xe = x$ in $S$ for some idempotent $e \in E(S)$. Write $w = x w'$. We have to distinguish two cases. The first case is $e \notin E(w')$. By induction, there exists a factorization $w' = x_1 w_1 y_1 \cdots x_m w_m y_m s$ with $m \leq |E(w')| < |E(w)|$ and $|x_1 y_1 \cdots x_m y_m s| \leq 2|S| |E(w')| + |S|$ satisfying conditions "1" and "3". If $m \geq 1$, then we set $x_1' = x x_1$. Now, $w = x_1' w_1 y_1 \cdots x_m w_m y_m s$ is a desired factorization of $w$. If $m = 0$, then the factorization for $w$ is $w = s'$ with $m = 0$.

The second case is $e \in E(w')$. Let $w' = w_0 y_0 w''$ such that $y_0 \in S^+$, $|y_0| \leq |S|$, $y_0 e = y_0$ in $S$ and $e \notin E(w'')$, i.e., we take $y_0$ as the last short factor of $w'$ such that it is stabilized by $e$. By induction, there exists a factorization $w'' = x_1 w_1 y_1 \cdots x_m w_m y_m s$. Now, $w = x_0 w_0 y_0 \cdots x_m w_m y_m s$ with $x_0 = x$ is a factorization of $w$ of the desired form. □

**Lemma 8.** *Let $S \in \mathbf{LR}$ be a finite semigroup. Let $u, x \in S^+$ and $e \in S$ be idempotent such that $u = ue$ and $x = xe$ in $S$. If $ux \mathrel{\mathcal{R}} u$, then $ux = u$ in $S$.*

*Proof.* Let $y \in S^*$ such that $uxy = u$. In $S$ we have $u = u(exeye)^\omega = u(exeye)^\omega exe = ux$ where the second equality follows because $S \in \mathbf{LR}$. □

**Proposition 9.** *Let $L \subseteq \Gamma^+$ be recognized by $S \in [\![x^\omega y x^\omega \leq x^\omega]\!]$. Then $L$ is a union of monomials $w_1 \Gamma^* w_2 \cdots \Gamma^* w_n$ of degree $|w_1 w_2 \cdots w_n| < 2|S|^3 + |S|^2$ and $n \leq |S|^2$.*

*Proof.* Let $h : \Gamma^+ \to S$ be a homomorphism recognizing $L$. We define the *depth* of the word $u \in \Gamma^+$ as $d(u) = |\{s \in S \mid h(u) \leq_\mathcal{R} s\}|$. For each $u \in \Gamma^+$ we construct a language $P_u = w_1 \Gamma^* w_2 \cdots \Gamma^* w_n$ with $|w_1 w_2 \cdots w_n| < 2d(u)|S|^2 + d(u)|S|$ such that
$$u \in P_u \subseteq h^{-1}(\downarrow h(u)).$$

In order to avoid unnecessary case distinctions, we set $P_\varepsilon = \varepsilon$ and $h(\varepsilon) >_\mathcal{R} h(u)$ for all $u \in \Gamma^+$. Let $u = vw$, $v \in \Gamma^*$, $w \in a\Gamma^*$ such that $h(v) >_\mathcal{R} h(va) \mathrel{\mathcal{R}} h(u)$. Now, $d(v) < d(u)$ and hence by induction, there exists a monomial $P_v$ with $v \in P_v \subseteq h^{-1}(\downarrow h(v))$ of degree less than $2d(u)|S|^2 + d(u)|S| - 2|S|^2 - |S|$. By Lemma 7 we find a factorization $w = x_1 u_1 y_1 \cdots x_m u_m y_m s$ such that $|x_1 y_1 \cdots x_m y_m s| < 2|S|^2 + |S|$ and for all $i \in \{1, \ldots, m\}$ there exists an idempotent $e_i$ with $h(x_i)e_i = h(x_i)$ and $h(y_i)e_i = h(y_i)$. Using Lemma 8 we see $h(u) = h(vw) = h(vx_1 \cdots x_m s)$. Now, define the monomial $P_u = P_v x_1 \Gamma^* y_1 x_2 \Gamma^* \cdots y_{m-1} x_m \Gamma^* y_m s$ of degree less than $2d(u)|S|^2 + d(u)|S|$. By construction $u \in P_u$. Consider $v'w' \in P_u$ with $v' \in P_v$ and $w' = x_1 w'_1 y_1 x_2 w'_2 \cdots y_{m-1} x_m w'_m y_m s$. We have $h(v') \leq h(v)$ and since $ese \leq e$ for all $s \in S$ and all $e \in E(S)$ we see that $h(x_i) = h(x_i)e_i \geq h(x_i)e_i h(w'_i y_i)e_i = h(x_i w'_i y_i)$. Therefore, $h(x_1 \cdots x_m s) \geq h(w')$ and $h(u) = h(vx_1 \cdots x_m s) \geq h(v'w')$.

Now with the above properties, $L \subseteq \bigcup_{u \in L} P_u \subseteq L$ and this union is finite since there are only finite many monomials of degree less than $2|S|^3 + |S|^2$. □

We are now ready to prove Theorem 3.

*Proof (Theorem 3).* "1 ⇒ 2": This is Lemma 5. "2 ⇒ 1" follows from property "1" of Lemma 4 and the fact that $\Sigma_1[<, +1, \min, \max]$ is closed under disjunction.

The implication "2 ⇒ 3" is trivial, and "3 ⇒ 4" is Lemma 6 since the class of languages recognizable by semigroups in $[\![x^\omega y x^\omega \leq x^\omega]\!]$ is closed under positive Boolean combinations. "4 ⇒ 5" is trivial. Finally, "5 ⇒ 2" follows immediately from Proposition 9. □

## 4 Existential First-order Logic without min or max

At higher levels of the quantifier alternation hierarchy, it is possible to specify the prefix and the suffix of a word by using successor $+1$ as the only predicate (apart from labels $\lambda(x) = a$ for $a \in \Gamma$). At the level $\Sigma_1$, the min-predicate is required to determine prefixes, and max is required for suffixes. We have the following inclusions:

$$\Sigma_1[<] \subsetneq \Sigma_1[<, +1] \begin{array}{c} \subsetneq \Sigma_1[<, +1, \min] \subsetneq \\ \subseteq \Sigma_1[<, +1, \max] \subsetneq \end{array} \Sigma_1[<, +1, \min, \max]$$

By a result of Pin [10], it is decidable whether a given regular language is definable in $\Sigma_1[<]$. For $\Sigma_1[<,+1,\min,\max]$, decidability follows by a result of Pin and Weil [11] (or alternatively by Theorem 3). In this section, we characterize the languages definable in the other fragments and we show that definability within these fragments is decidable. As it turns out, these decidability results are a combination of effective algebraic and effective topological properties, cf. [7].

**Theorem 10.** *Let $L \subseteq \Gamma^+$. The following assertions are equivalent:*
1. *$L$ is definable in $\Sigma_1[<,+1,\min]$.*
2. *$L$ is a finite union of monomials $w_1 \Gamma^* \cdots w_n \Gamma^*$.*
3. *$\mathrm{Synt}(L) \in [\![x^\omega y x^\omega \leq x^\omega]\!]$ and $h_L(L)$ is a $\leq_\mathcal{R}$-order ideal.*

*Proof.* "1 ⇒ 2": Let $L = L(\varphi)$ for $\varphi \in \Sigma_1[<,+1,\min]$. By Theorem 3, the language $L$ is a finite union of monomials $w_1 \Gamma^* w_2 \cdots \Gamma^* w_n$. Let $u \models \varphi$. Then for every $v \in \Gamma^*$ the same assignment of the variables which makes $\varphi$ true on $u$ also satisfies $\varphi$ on $uv$. Therefore, $L\Gamma^* \subseteq L$. Since $(P \cup Q)\Gamma^* = P\Gamma^* \cup Q\Gamma^*$, it follows that $L$ is a finite union of monomials $w_1 \Gamma^* w_2 \cdots \Gamma^* w_n \Gamma^*$.

"2 ⇒ 1": This follows from "2" in Lemma 4.

"2 ⇒ 3": $\mathrm{Synt}(L) \in [\![x^\omega y x^\omega \leq x^\omega]\!]$ follows from Theorem 3. By [7, Theorem 1] we see that $h_L(L)$ is a $\leq_\mathcal{R}$-order ideal. The implication "3 ⇒ 2" follows from Proposition 9 and [7, Theorem 1]. □

Of course, there also is a left-right dual of the above theorem: A language $L$ is definable in $\Sigma_1[<,+1,\max]$ if and only if $L$ is a union of monomials of the form $\Gamma^* w_1 \cdots \Gamma^* w_n$ if and only if $\mathrm{Synt}(L) \in [\![x^\omega y x^\omega \leq x^\omega]\!]$ and $h_L(L)$ is a $\leq_\mathcal{L}$-order ideal. The following theorem is the analogue of Theorem 10 with neither min nor max predicates.

**Theorem 11.** *Let $L \subseteq \Gamma^+$. The following assertions are equivalent:*
1. *$L$ is definable in $\Sigma_1[<,+1]$.*
2. *$L$ is a finite union of monomials $\Gamma^* w_1 \cdots \Gamma^* w_n \Gamma^*$.*
3. *$\mathrm{Synt}(L) \in [\![x^\omega y x^\omega \leq x^\omega]\!]$ and $h_L(L)$ is a $\leq_\mathcal{J}$-order ideal.*

*Proof.* "1 ⇒ 2": Let $L$ be defined by $\varphi \in \Sigma_1[<,+1]$. By Theorem 3, $L$ is a finite union of monomials $w_1 \Gamma^* w_2 \cdots \Gamma^* w_n$. Let $u,v,w \in \Gamma^*$. Every assignment satisfying $\varphi$ on $u$ also satisfies $\varphi$ on $vuw$. Hence, $\Gamma^* L \Gamma^* \subseteq L$.

"2 ⇒ 1": This follows from "3" in Lemma 4.

"2 ⇒ 3": $\mathrm{Synt}(L) \in [\![x^\omega y x^\omega \leq x^\omega]\!]$ follows from Theorem 3; the set $h_L(L)$ is a $\leq_\mathcal{J}$-order ideal by [7, Theorem 3]. The implication "3 ⇒ 2" follows from Proposition 9 and [7, Theorem 3]. □

The following decidability result is an immediate consequence of our characterizations.

**Corollary 12.** *Let $L \subseteq \Gamma^+$ be a regular language. It is decidable whether $L$ is definable in $\Sigma_1[<,+1]$ (resp. $\Sigma_1[<,+1,\min]$, resp. $\Sigma_1[<,+1,\max]$).*

*Proof.* The syntactic homomorphism $h_L : \Gamma^+ \to \mathrm{Synt}(L)$ of $L$ is effectively computable. Hence, one can verify whether property "3" in Theorem 11 (resp. "3" in Theorem 10, resp. the left-right dual of "3" in Theorem 10) holds. □

## 5 Dot-Depth One

A language $L \subseteq \Gamma^+$ has dot-depth 1 if it is a Boolean combination of monomials of the form $w_1\Gamma^*w_2\cdots\Gamma^*w_n$ with $w_i \in \Gamma^*$. Knast [6] has shown that a language $L$ has dot-depth 1 if and only if $\mathrm{Synt}(L) \in \mathbf{B_1}$. Since the latter property is decidable, this gives decidability of dot-depth 1. Later, Thérien [19] gave a simpler proof for Knast's result. Both proofs are based on an algebraic concept called *finite categories*, see [21]. In this section, we give a new (more combinatorial) proof of this theorem. As for dot-depth 1/2, the main advantage of our proof is that the bounds involved are more explicit.

**Theorem 13** (Thomas [20], Knast [6])**.** *Let $L \subseteq \Gamma^+$. The following assertions are equivalent:*
1. *$L$ is definable in $\mathbb{B}\Sigma_1[<, +1, \min, \max]$.*
2. *$L$ is a Boolean combination of monomials $w_1\Gamma^*w_2\cdots\Gamma^*w_n$.*
3. *$\mathrm{Synt}(L) \in \mathbf{B_1}$.*
4. *$L$ is recognized by some semigroup in $\mathbf{B_1}$.*

As for dot-depth 1/2, the equivalence of $\mathbb{B}\Sigma_1[<, +1, \min, \max]$ and dot-depth 1 is due to a result by Thomas [20]. The remainder of this section is devoted to the proof of the above theorem.

**Lemma 14.** *Let $S$ be a finite semigroup and let $u \in S^+$. Suppose there exists $e \in S$ such that $pe = p$ in $S$ for some prefix $p \leq_p u$. Choose $|p|$ maximal with this property and let $u = pv$. If $xpvy = x'pvy'$ for some $x \neq x'$, then there is at least one letter between the factors $v$ in the two factorizations.*

*Proof.* Let $|x'| < |x|$ and assume that the claim is not true.

| $x$ | | $p$ | | $v$ | | $y$ |
|---|---|---|---|---|---|---|
| $x'$ | | $p$ | | $v$ | | $y'$ |
| | | $p'$ | | | $v'$ | |

Then we find a factorization $pv = p'v'$ such that $p'e = p'$ and $|p'| > |p|$ contradicting the maximality of $|p|$. This also holds if the factors do not overlap but are adjacent, in which case $p' = pv$. □

The following lemma will serve as the link between the algebraic properties of $\mathbf{B_1}$ and the combinatorial properties in Lemma 16 below.

**Lemma 15.** *Let $S \in \mathbf{LR}$ and let $k \geq |S| + 1$. For every $a \in S$ and for all $u, x \in S^+$ with $|x| \geq k$ we have:* $u \mathcal{R} ux >_\mathcal{R} uxa \Rightarrow \mathrm{alph}_k(x) \neq \mathrm{alph}_k(xa)$.

*Proof.* Suppose $u \mathcal{R} ux$ and $\mathrm{alph}_k(x) = \mathrm{alph}_k(xa)$. Let $w$ be the suffix of $xa$ of length $k$. By Lemma 1, there exist $p, v \in S^*$ such that $w = pva$ in $S^+$ and $pe = p$ in $S$ for some idempotent $e \in S$. Let $|p| \leq |S|$ be maximal with this property. Since $w \in \mathrm{alph}_k(xa) = \mathrm{alph}_k(x)$ we can write
$$x = spvatv \text{ in } S^+$$
for some $s, t \in S^*$ such that $p$ is a suffix of $pvat$. Note that there is indeed at least one letter between the two occurrences of $v$ by Lemma 14. For $u' = usp$ and $x' = vat$ we have $u' = u'e$, $u'x' = u'x'e$, and $u' \mathcal{R} u'x'$. Using Lemma 8 we see that $u' = u'x' = u'x'x'$. Hence, $u \mathcal{R} u' = u'x'x' = uxat$ and therefore, $u \mathcal{R} uxa$. □

The following lemma is the main combinatorial ingredient for our proof of Knast's Theorem. It generalizes an idea of Klíma [5] to factors of words. The determinacy mechanism is similar to unary interval logic with lookaround [8].

**Lemma 16.** *Let $x_i, y_i, u_i', u_i', v_i, v_i' \in \Gamma^+$ and $u_k, v_k, u_1', v_1' \in \Gamma^*$, and let*

$$u = x_1 u_1 \cdots x_k u_k = u_1' y_1 \cdots u_\ell' y_\ell$$
$$v = x_1 v_1 \cdots x_k v_k = v_1' y_1 \cdots v_\ell' y_\ell$$

*such that $x_1 u_1 \cdots x_k$ (resp. $x_1 v_1 \cdots x_k$) is the shortest prefix of $u$ (resp. $v$) in $x_1 \Gamma^+ x_2 \cdots \Gamma^+ x_k$, and $y_1 \cdots u_\ell' y_\ell$ (resp. $y_1 \cdots v_\ell' y_\ell$) is the shortest suffix of $u$ (resp. $v$) in $y_1 \Gamma^+ y_2 \cdots \Gamma^+ y_\ell$.*

*If $u$ and $v$ are contained in the same monomials $w_1 \Gamma^+ w_2 \cdots \Gamma^+ w_n$ with $n \leq k + \ell$ and degree $|w_1 \cdots w_n| \leq |x_1 \cdots x_k y_1 \cdots y_\ell|$, then the relative positions of $x_k$ and $y_1$ are the same in $u$ as in $v$. More precisely,*

1. *$x_1 u_1 \cdots x_k$ is a prefix of $u_1'$ iff $x_1 v_1 \cdots x_k$ is a prefix of $v_1'$,*
2. *if $x_k$ and $y_1$ overlap in $u$ or in $v$, then they have the same overlap in both words,*
3. *$u_1' y_1$ is a prefix of $x_1 \cdots u_{k-1}$ iff $v_1' y_1$ is a prefix of $x_1 \cdots v_{k-1}$.*

*Proof.* "1": Suppose that $x_1 u_1 \cdots x_k$ is a prefix of $u_1'$. Then $u$ is contained in the language $x_1 \Gamma^+ \cdots x_k \Gamma^+ v_1 \cdots \Gamma^+ v_\ell$ or in $x_1 \Gamma^+ \cdots x_k v_1 \cdots \Gamma^+ v_\ell$. Hence $v$ is contained in one of these two monomials, showing that $x_1 v_1 \cdots x_k$ is a prefix of $v_1'$.

"2": We can assume that none of the conditions in "1" holds. We have to distinguish two cases. First, suppose that $x_k$ and $y_1$ overlap in $u$ such that $x_1 u_1 \cdots x_k$ is a prefix of $u_1' y_1$ and let $z$ be the word comprising all positions of $x_k$ and $y_1$ in $u$. Then $u \in P = x_1 \Gamma^+ \cdots x_{k-1} \Gamma^+ z \Gamma^+ y_2 \cdots \Gamma^+ v_\ell$. Hence $v \in P$, showing that $x_k$ and $y_1$ in $v$ have at most the same overlap as in $u$.

The second case is that $x_k$ and $y_1$ overlap in $u$ such that $x_1 u_1 \cdots x_k$ is not a prefix of $u_1' y_1$. Moreover, we can assume that $x_1 v_1 \cdots x_k$ is not a prefix of $v_1' y_1$ since otherwise we are in the first case with $u$ and $v$ interchanged. Now, $u$ is contained in $P = x_1 \Gamma^+ \cdots x_i \Gamma^+ z \Gamma^+ y_j \cdots \Gamma^+ v_\ell$, where $z$ is the factor of $u$ comprising all $x_{i+1}, \ldots, x_k$ which are overlapping (or adjacent) with $y_1$ and all $y_1, \ldots, y_{j-1}$ which are overlapping (or adjacent) with $x_k$. Since $v \in P$, we conclude that $x_k$ and $y_1$ in $v$ have at least the same overlap as in $u$.

"3": If none of the conditions in "1" and "2" holds, then in both words $u$ and $v$, the factor $y_1$ is on the left-hand side of $x_k$. □

**Lemma 17.** *Let $S \in \mathbf{B_1}$. For all $u, v, x, s \in S$ and for all $e, f \in E(S)$, the following implication holds: $u \mathrel{\mathcal{R}} uexf$, $esfv \mathrel{\mathcal{L}} v \Rightarrow uexfv = uesfv$.*

*Proof.* Since $u \mathrel{\mathcal{R}} uexf$ and $v \mathrel{\mathcal{L}} esfv$, there exist $y, t \in M$ with $u = uexfy$ and $v = tesfv$. In particular, $u = u(exfy)^\omega$ and $v = (tesf)^\omega v$. We conclude

$$uexfv = u(exfy)^\omega exf(tesf)^\omega v = u(exfy)^\omega esf(tesf)^\omega v = uesfv,$$

where the second equality uses $S \in \mathbf{B_1}$. □

**Proposition 18.** *Let $L \subseteq \Gamma^+$ be recognized by $h : \Gamma^+ \to S$ with $S \in \mathbf{B_1}$. If words $u$ and $v$ are contained in the same monomials $w_1 \Gamma^+ w_2 \cdots \Gamma^+ w_n$ with $n \leq 2|S|$ and degree $|w_1 \cdots w_n| \leq 4|S|^2 - 2|S|$, then $h(u) = h(v)$.*

*Proof.* This proof was inspired by Klíma's proof [5] of Simon's Theorem on piecewise-testable languages. The outline of our proof is as follows. We consider factorizations induced by the $\mathcal{R}$-factorization of $u$ and the $\mathcal{L}$-factorization of $v$. Then we transfer the factorization of $u$ to $v$ and vice versa such that the respective orders of the factor in $u$ and $v$ are the same. Finally, we transform $u$ into $v$ by a sequence of $h$-invariant substitutions.

Consider the $\mathcal{R}$-factorization $u = a_1 u_1 \cdots a_k u_k$ such that

$$h(a_1 u_1 \cdots a_i) \mathrel{\mathcal{R}} h(a_1 u_1 \cdots a_i u_i) >_{\mathcal{R}} h(a_1 u_1 \cdots a_i u_i a_{i+1})$$

for all $i$. We have $k \leq |S|$. Let $j_i$ be the position of $a_i$ in the above factorization. We color all positions of $u$ in any of the intervals $[\,j_i - |S|\,;\, j_i + |S| - 1\,]$ red. In particular, the $a_i$-positions $j_i$ are red. And in general, there is a neighborhood of size $2|S|$ around each $a_i$ which contains only red positions. In the worst case, $a_1$ is the only exception. Hence, there are at most $2|S|^2 - |S|$ red positions in $u$. Let $R_i$ be the $i$-th consecutive factor of red positions. Then $u = R_1 u'_1 \cdots R_{k'} u'_{k'}$ for some $u'_i \in \Gamma^+$, $i < k'$, and $u'_{k'} \in \Gamma^*$. Note that $k' \leq k$ since some intervals could overlap. By Lemma 15, the word $R_1 u'_1 \cdots R_i$ is the shortest prefix of $u$ contained in $R_1 \Gamma^+ \cdots R_i$.

Symmetrically, we consider the $\mathcal{L}$-factorization $v = v_1 b_1 \cdots v_\ell b_\ell$ such that

$$h(b_{i+1} v_i b_i \cdots v_1 b_1) <_{\mathcal{L}} h(v_i b_i \cdots v_1 b_1) \mathrel{\mathcal{L}} h(b_i \cdots v_1 b_1)$$

for all $i$. Let $j'_i$ be the position of $b_i$ in the above factorization. We color all positions of $v$ in any of the intervals $[\,j'_i - |S| + 1\,;\, j'_i + |S|\,]$ blue. As before, there are at most $2|S|^2 - |S|$ blue positions. Let $B_i$ be the $i$-th consecutive factor of blue positions. Then $v = v'_1 B_1 \cdots v'_{\ell'} B_{\ell'}$ for some $v'_i \in \Gamma^+$, $i \geq 1$ and $v'_1 \in \Gamma^*$. As before, $B_i \cdots v'_{\ell'} B_{\ell'}$ is the shortest suffix of $v$ contained in $B_i \cdots \Gamma^+ B_{\ell'}$.

Next, we transfer the red positions of $u$ to $v$, and we transfer the blue positions of $v$ to $u$. By assumption, $v \in R_1 \Gamma^+ \cdots R_{k'} \Gamma^+$. Therefore, there exists a factorization $v = R_1 v''_1 \cdots R_{k'} v''_{k'}$ such that $R_1 v''_1 \cdots R_i$ is the shortest prefix of $v$ contained in $R_1 \Gamma^+ \cdots R_i$. We color the positions of the $R_i$'s in $v$ red. Similarly, there exists a factorization $u = u''_1 B_1 \cdots u''_{\ell'} B_{\ell'}$ such that $B_i \cdots u''_{\ell'} B_{\ell'}$ is the shortest suffix of $u$ contained in $B_i \cdots \Gamma^+ B_{\ell'}$. We color the positions of the $B_i$'s in $u$ blue. Now, colored positions in $u$ and $v$ are either red or blue or both. By Lemma 16, the colored positions in $u$ have the same order as the colored positions in $v$. Let $w_i$ be the $i$-th consecutive factor of colored (red or blue) positions, and write

$$u = w_1 x_1 \cdots w_{n-1} x_{n-1} w_n,$$
$$v = w_1 s_1 \cdots w_{n-1} s_{n-1} w_n.$$

By Lemma 1 and its left-right dual, there exist $e_1, \ldots, e_{n-1} \in E(S)$ and $f_2, \ldots, f_n \in E(S)$ such that each $w_i$ admits a factorization $w_i = p_i r_i q_i$ with $|p_i| \leq |S| - 1$ and $|q_i| \leq |S| - 1$ satisfying

$$h(r_i) = h(r_i)\, e_i \quad \text{for } 1 \leq i < n,$$
$$h(r_i) = f_i\, h(r_i) \quad \text{for } 1 < i \leq n.$$

In particular, we can assume $p_1 = \varepsilon = q_n$. Let $x'_i = q_i x_i p_{i+1}$ and $s'_i = q_i s_i p_{i+1}$ for $1 \leq i < n$. Then

$$u = r_1 x'_1 r_2 \cdots x'_{n-1} r_n,$$
$$v = r_1 s'_1 r_2 \cdots s'_{n-1} r_n,$$

and the $r_i$'s in $u$ cover the positions of the $\mathcal{R}$-factorization of $u$, whereas the $r_i$'s in $v$ cover the positions of the $\mathcal{L}$-factorization of $v$. Therefore, we have

$$h(r_1 x'_1 \cdots r_i) \;\; \mathcal{R} \;\; h(r_1 x'_1 \cdots r_i) \cdot e_i h(x'_i) f_{i+1} \qquad \text{for all } 1 \leq i < n,$$
$$h(r_i \cdots s'_n r_n) \;\; \mathcal{L} \;\; e_{i-1} h(s'_{i-1}) f_i \cdot h(r_i \cdots s'_n r_n) \qquad \text{for all } 1 < i \leq n.$$

By an $(n-1)$-fold application of Lemma 17 we obtain

$$\begin{aligned} h(u) &= h(r_1 x'_1 \cdots r_{n-2} x'_{n-2} r_{n-1} x'_{n-1} r_n) \\ &= h(r_1 x'_1 \cdots r_{n-2} x'_{n-2} r_{n-1} s'_{n-1} r_n) \\ &= h(r_1 x'_1 \cdots r_{n-2} s'_{n-2} r_{n-1} s'_{n-1} r_n) \\ &= \cdots \\ &= h(r_1 s'_1 \cdots r_{n-2} s'_{n-2} r_{n-1} s'_{n-1} r_n) = h(v). \end{aligned}$$

Note that the substitution rules $x'_i \to s'_i$ are $h$-invariant in their respective contexts only when applied from right to left when converting $h(u)$ into $h(v)$. □

**Corollary 19.** *Let $L \subseteq \Gamma^+$ be recognized by a finite semigroup $S \in \mathbf{B_1}$. If words $u$ and $v$ are contained in the same monomials $w_1 \Gamma^* w_2 \cdots \Gamma^* w_n$ with $n \leq 2|S|$ and degree $|w_1 \cdots w_n| < 4|S|^2$, then $h(u) = h(v)$.*

*Proof.* Every monomial $w_1 \Gamma^+ \cdots w_{n-1} \Gamma^+ w_n$ is a union of monomials of the form

$$w_1 a_1 \Gamma^* \cdots w_{n-1} a_{n-1} \Gamma^* w_n$$

for $a_1, \ldots, a_{n-1} \in \Gamma$. Therefore, the claim follows from Proposition 18. □

We are now ready to prove Theorem 13.

*Proof (Theorem 13).* "1 ⇔ 2": This follows from Theorem 3.

"2 ⇒ 3": By Lemma 6 the syntactic semigroup of every monomial $w_1 \Gamma^* w_2 \cdots \Gamma^* w_n$ satisfies $x^\omega y x^\omega \leq x^\omega$ and by Lemma 2 it is in $\mathbf{B_1}$. The claim follows since the class of languages recognizable in $\mathbf{B_1}$ is closed under Boolean combinations. The implication "3 ⇒ 4" is trivial.

"4 ⇒ 2": Let $L$ be recognized by $h : \Gamma^+ \to S \in \mathbf{B_1}$. We write $u \equiv v$ if $u$ and $v$ are contained in the same monomials of the form $w_1 \Gamma^* w_2 \cdots \Gamma^* w_n$ of degree at most $4|S|^2$. We have $L = h^{-1}(P)$ for $P = h(L)$. Corollary 19 shows that every set $h^{-1}(p)$ is a union of $\equiv$-classes. Moreover, $\equiv$ has finite index since there are only finitely many monomials of bounded degree. Every $\equiv$-class is a finite Boolean combination of the required form by specifying which monomials hold and which do not. □

## 6 Dot-depth One without min or max

As for $\Sigma_1$, one cannot define min- or max-predicates in $\mathbb{B}\Sigma_1[<, +1]$. Therefore, the following inclusions hold:

$$\mathbb{B}\Sigma_1[<] \;\subsetneq\; \mathbb{B}\Sigma_1[<, +1] \;\begin{array}{c} \subsetneq \\ \\ \subsetneq \end{array}\; \begin{array}{c} \mathbb{B}\Sigma_1[<, +1, \min] \\ \\ \mathbb{B}\Sigma_1[<, +1, \max] \end{array}\; \begin{array}{c} \subsetneq \\ \\ \subsetneq \end{array}\; \mathbb{B}\Sigma_1[<, +1, \min, \max]$$

Simon's Theorem on piecewise testable languages [14] gives decidability of $\mathbb{B}\Sigma_1[<]$. For the fragment $\mathbb{B}\Sigma_1[<,+1,\min,\max]$, decidability follows by Knast's Theorem [6], see Theorem 13. In this section, we give effective characterizations of the remaining fragments. As for dot-depth $1/2$, these characterizations are a combination of algebraic and topological properties, cf. [7]. Moreover, we obtain natural subclasses of dot-depth 1 for the languages definable by the above fragments.

**Lemma 20.** *Let $P = w_1\Gamma^*w_2\cdots\Gamma^*w_n$ and let $uq \in P$. Then there exists a monomial $P' = v_1\Gamma^*v_2\cdots\Gamma^*v_n$ with $|v_1\cdots v_n| \leq |w_1\cdots w_n q|$ such that $uq \in P' \subseteq (Pq^{-1})q$.*

*Proof.* Let $j \leq n$ be minimal such that $q \in \Gamma^*w_j\cdots\Gamma^*w_n$. If there exists a proper prefix $y$ of $w_{j-1}$ such that $y$ is a suffix of $u$ and $yq \in w_{j-1}\cdots\Gamma^*w_n$, then we set $q' = yq$, else we set $q' = q$. We assume $q'$ to be maximal with these properties. We can write $uq = u''q'$. Moreover, by maximality of $q'$, there exists an index $j'$ such that $q' \in \Gamma^*w_{j'}\cdots\Gamma^*w_n$ and $q' \notin y'\Gamma^*w_{j'}\cdots\Gamma^*w_n$ for any non-empty suffix $y'$ of $w_{j'-1}$. We set $P' = w_1\Gamma^*\cdots w_{j'-1}\Gamma^*q'$. Now, $uq \in P'$ and for all $w \in P'$ we have $w \in P \cap \Gamma^*q = (Pq^{-1})q$. □

**Lemma 21.** *Let $h : \Gamma^+ \to S \in \mathbf{B_1}$. If $u,v \in \Gamma^+$ are contained in the same monomials $w_1\Gamma^*\cdots w_n\Gamma^*$ of degree $|w_1\cdots w_n| < 8|S|^2$, then $h(u)\,\mathcal{R}\,h(v)$.*

*Proof.* We write $u \equiv_m v$, if $u$ and $v$ are contained in the same monomials $w_1\Gamma^*w_2\cdots\Gamma^*w_n$ of degree $|w_1\cdots w_n| \leq m$. Analogously, we write $u \sim_m v$ if $u$ and $v$ are contained in the same monomials $w_1\Gamma^*\cdots w_n\Gamma^*$ of degree $|w_1\cdots w_n| \leq m$. If $u \equiv_m v$ for $m = 4|S|^2 - 1$, then by Corollary 19 we have $h(u) = h(v)$.

Let $u \sim_{2m} v$. We want to show $h(u)\,\mathcal{R}\,h(v)$. We can assume $|u|,|v| \geq 2m$, because otherwise $u = v$. Let $u = u'q$ with $|q| = m$. Consider the factorization $v = v'qx$ such that $qx$ is the shortest suffix of $v$ admitting $q$ as a factor, i.e., $v$ is factorized at the last occurrence of $q$. This factorization exists, since $u \in \Gamma^*q\Gamma^* \ni v$. We claim $u \equiv_k v'q$ and therefore, $h(v) \leq_{\mathcal{R}} h(v'q) = h(u)$. Symmetry then yields $h(u)\,\mathcal{R}\,h(v)$.

We now prove the claim. First, let $v'q \in P = w_1\Gamma^*w_2\cdots\Gamma^*w_n$ with $w_1\cdots w_n \leq m$. Then $v \in P\Gamma^*$ and $u \in P\Gamma^*$. Since $w_n$ is a suffix of $q$, we conclude $u \in P$.

Next, suppose $u \in P = w_1\Gamma^*w_2\cdots\Gamma^*w_n$ with $|w_1\cdots w_n| \leq m$. By Lemma 20, there exists a monomial $P' = v_1\Gamma^*v_2\cdots\Gamma^*v_n$ with $|v_1\cdots v_n| \leq |w_1\cdots w_n q| \leq 2m$ and $u'q \in P' \subseteq (Pq^{-1})q$. Since $u \in P'\Gamma^*$, we obtain $v \in P'\Gamma^*$. By choice of $x$, we have $v'q \in P'\Gamma^* \subseteq P\Gamma^*$. Since $w_n$ is a suffix of $q$, we conclude $v'q \in w_1\Gamma^*w_2\cdots\Gamma^*w_n$. □

**Theorem 22.** *Let $L \subseteq \Gamma^+$. The following assertions are equivalent:*
1. *$L$ is definable in $\mathbb{B}\Sigma_1[<,+1,\min]$.*
2. *$L$ is a Boolean combination of monomials $w_1\Gamma^*\cdots w_n\Gamma^*$.*
3. *$\mathrm{Synt}(L) \in \mathbf{B_1}$ and the syntactic homomorphism $h_L : \Gamma^+ \to \mathrm{Synt}(L)$ has the property that $h_L(L)$ is a union of $\mathcal{R}$-classes.*

*Proof.* The equivalence "1 ⇔ 2" follows from Theorem 10.

"2 ⇒ 3": We have $\mathrm{Synt}(L) \in \mathbf{B_1}$ by Theorem 13, and $h_L(L)$ is a union of $\mathcal{R}$-classes by [7, Theorem 5].

"3 ⇒ 2": By Lemma 21, there exists $m \in \mathbb{N}$ such that $h_L(u)\,\mathcal{R}\,h_L(v)$ if $u$ and $v$ are contained in the same languages of the form $w_1\Gamma^*\cdots w_n\Gamma^*$ with $|w_1\cdots w_n| \leq m$. Therefore, for each $\mathcal{R}$-class $R$ of $\mathrm{Synt}_L(L)$, the language $h_L^{-1}(R)$ is a Boolean combination of languages $w_1\Gamma^*\cdots w_n\Gamma^*$ with $|w_1\cdots w_n| \leq m$. The claim follows, since $L$ is a union of languages of the form $h_L^{-1}(R)$. □

There also is a left-right dual of the above theorem: A language $L$ is definable in $\mathbb{B}\Sigma_1[<,+1,\max]$ if and only if $L$ is a Boolean combination of monomials $\Gamma^*w_1 \cdots \Gamma^*w_n$ if and only if $\text{Synt}(L) \in \mathbf{B_1}$ and $h_L(L)$ is a union of $\mathcal{L}$-classes. Next, we consider the fragment $\mathbb{B}\Sigma_1[<,+1]$ with neither min nor max.

**Lemma 23.** *Let $h : \Gamma^+ \to S \in \mathbf{B_1}$. If $u, v \in \Gamma^+$ are contained in the same monomials $\Gamma^*w_1\Gamma^* \cdots w_n\Gamma^*$ of degree $|w_1 \cdots w_n| < 12|S|^2$, then $h(u) \mathcal{J} h(v)$.*

*Proof.* This proof is only a slight variation of the proof of Lemma 21. We write $u \equiv_m v$, if $u$ and $v$ are contained in the same monomials $w_1\Gamma^* \cdots \Gamma^*w_n$ of degree $|w_1 \cdots w_n| \leq m$. Analogously, we write $u \sim_m v$, if $u$ and $v$ are contained in the same monomials $\Gamma^*w_1\Gamma^* \cdots \Gamma^*w_n\Gamma^*$ of degree $|w_1 \cdots w_n| \leq m$. If $u \equiv_m v$ for $m = 4|S|^2 - 1$, then by Corollary 19 we have $h(u) = h(v)$.

Let $u \sim_{3m} v$. We want to show $h(u) \mathcal{J} h(v)$. We can assume $|u|, |v| \geq 3m$, because otherwise $u = v$. Let $u = pu'q$ with $|p| = |q| = m$. Consider the factorization $v = spv'qx$ such that $sp$ is the shortest prefix of $v$ admitting $p$ as a factor and $qx$ is the shortest suffix of $v$ admitting $q$ as a factor, i.e., $v$ is factorized at the first occurrence of $p$ and the last occurrence of $q$. This factorization exists, since $u \in \Gamma^*p\Gamma^*q\Gamma^* \ni v$. We claim $u \equiv_m pv'q$ and therefore, $h(v) \leq_{\mathcal{J}} h(pv'q) = h(u)$. Symmetry then yields $h(u) \mathcal{J} h(v)$.

We now prove the claim. First, let $pv'q \in P$ for $P = w_1\Gamma^*w_2 \cdots \Gamma^*w_n$ with $w_1 \cdots w_n \leq m$. Then $v \in \Gamma^*P\Gamma^*$ and $u \in \Gamma^*P\Gamma^*$. Since $w_1$ is a prefix of $p$ and $w_n$ is a suffix of $q$, we conclude $u \in P$.

Next, suppose $u \in P$ with $|w_1 \cdots w_n| \leq m$. By Lemma 20 and its left-right dual, there exists a monomial $P' = v_1\Gamma^*v_2 \cdots \Gamma^*v_n$ with $|v_1 \cdots v_n| \leq |pw_1 \cdots w_nq| \leq 3m$ and $u = pu'q \in P' \subseteq p(p^{-1}Pq^{-1})q$. Since $u \in \Gamma^*P'\Gamma^*$, we obtain $v \in \Gamma^*P'\Gamma^*$. By choice of $s$ and $x$, we have $pv'q \in \Gamma^*P'\Gamma^* \subseteq \Gamma^*P\Gamma^*$. Since $w_1$ is a prefix of $p$ and $w_n$ is a suffix of $q$, we conclude $pv'q \in w_1\Gamma^*w_2 \cdots \Gamma^*w_n$. □

**Theorem 24.** *Let $L \subseteq \Gamma^+$. The following assertions are equivalent:*

1. *$L$ is definable in $\mathbb{B}\Sigma_1[<,+1]$.*
2. *$L$ is a Boolean combination of monomials $\Gamma^*w_1 \cdots \Gamma^*w_n\Gamma^*$.*
3. *$\text{Synt}(L) \in \mathbf{B_1}$ and the syntactic homomorphism $h_L : \Gamma^+ \to \text{Synt}(L)$ has the property that $h_L(L)$ is a union of $\mathcal{J}$-classes.*

*Proof.* The equivalence "1 $\Leftrightarrow$ 2" follows from Theorem 11.

"2 $\Rightarrow$ 3": We have $\text{Synt}(L) \in \mathbf{B_1}$ by Theorem 13, and $h_L(L)$ is a union of $\mathcal{R}$-classes by [7, Theorem 7].

"3 $\Rightarrow$ 2": By Lemma 23, there exists $m \in \mathbb{N}$ such that $h_L(u) \mathcal{J} h_L(v)$, if $u$ and $v$ are contained in the same languages of the form $\Gamma^*w_1\Gamma^* \cdots w_n\Gamma^*$ with $|w_1 \cdots w_n| \leq m$. Since $h_L(L)$ is a union of $\mathcal{J}$-classes, the language $L$ is a Boolean combination of languages of the form $\Gamma^*w_1 \cdots \Gamma^*w_n\Gamma^*$ of degree $|w_1 \cdots w_n| \leq m$. □

The following decidability result is an immediate consequence of our characterizations.

**Corollary 25.** *Let $L \subseteq \Gamma^+$ be a regular language. It is decidable whether $L$ is definable in $\mathbb{B}\Sigma_1[<,+1]$ (resp. $\mathbb{B}\Sigma_1[<,+1,\min]$, resp. $\mathbb{B}\Sigma_1[<,+1,\max]$).*

*Proof.* The syntactic homomorphism $h_L : \Gamma^+ \to \text{Synt}(L)$ of $L$ is effectively computable. Hence, one can verify whether property "3" in Theorem 24 (resp. "3" in Theorem 22, resp. the left-right dual of "3" in Theorem 22) holds. □

| Languages | Logics | Algebra | |
|---|---|---|---|
| $\bigcup w_1 \Gamma^* w_2 \cdots \Gamma^* w_n$ | $\Sigma_1[<, +1, \min, \max]$ | $\mathbf{B_{1/2}}$ | [11], Thm. 3 |
| $\bigcup w_1 \Gamma^* \cdots w_n \Gamma^*$ | $\Sigma_1[<, +1, \min]$ | $\leq_\mathcal{R}$-order ideals in $\mathbf{B_{1/2}}$ | Thm. 10 |
| $\bigcup \Gamma^* w_1 \cdots \Gamma^* w_n$ | $\Sigma_1[<, +1, \max]$ | $\leq_\mathcal{L}$-order ideals in $\mathbf{B_{1/2}}$ | left-right dual of Thm. 10 |
| $\bigcup \Gamma^* w_1 \cdots \Gamma^* w_n \Gamma^*$ | $\Sigma_1[<, +1]$ | $\leq_\mathcal{J}$-order ideals in $\mathbf{B_{1/2}}$ | Thm. 11 |
| $\mathbb{B}(w_1 \Gamma^* w_2 \cdots \Gamma^* w_n)$ | $\mathbb{B}\Sigma_1[<, +1, \min, \max]$ | $\mathbf{B_1}$ | [6], Thm. 13 |
| $\mathbb{B}(w_1 \Gamma^* \cdots w_n \Gamma^*)$ | $\mathbb{B}\Sigma_1[<, +1, \min]$ | $\mathcal{R}$-classes in $\mathbf{B_1}$ | Thm. 22 |
| $\mathbb{B}(\Gamma^* w_1 \cdots \Gamma^* w_n)$ | $\mathbb{B}\Sigma_1[<, +1, \max]$ | $\mathcal{L}$-classes in $\mathbf{B_1}$ | left-right dual of Thm. 22 |
| $\mathbb{B}(\Gamma^* w_1 \cdots \Gamma^* w_n \Gamma^*)$ | $\mathbb{B}\Sigma_1[<, +1]$ | $\mathcal{J}$-classes in $\mathbf{B_1}$ | Thm. 24 |

Table 1: Languages around dot-depth one.

## 7 Summary

We considered subclasses of languages with dot-depth 1/2 and of languages with dot-depth 1. These subclasses admit counterparts in terms of fragments of existential first-order logic $\Sigma_1$ and its Boolean closure $\mathbb{B}\Sigma_1$. For all fragments, we give effective algebraic characterizations. At closer look, the characterizations are a conjunction of an algebraic and a topological property, cf. [7]. We summarize our main results in Table 1. To shorten notation, we write $\mathbf{B_{1/2}}$ instead of $[\![x^\omega y x^\omega \leq x^\omega]\!]$.

In addition, we give new proofs for Pin and Weil's Theorem on dot-depth 1/2 and for Knast's Theorem on dot-depth 1. The proofs are combinatorial and they improve the bounds involved in computing a language description for a given recognizing semigroup.